\documentclass[reprint, amsmath, amssymb, aps]{revtex4-1}
\pdfoutput=1

\usepackage{graphicx}
\graphicspath{figures}
\usepackage{dcolumn}
\usepackage{bm}
\usepackage{hyperref}

\begin{document}
	\title{A Software Tool for "Gluing" Distributions}
	\author{Peter Werner}
	\email{peter.werner@uni-oldenburg.de}
	\affiliation{
		Institut f\"ur Physik, Universit\"at Oldenburg, 26111 Oldenburg, Germany
	}	
	\date{\today}
	\begin{abstract}
		When performing Monte-Carlo simulations, distributions are sometimes
		determined only for sub-intervals of the desired total range.
		In such cases, a frequent problem is to connect, or glue, individual 
		distributions to obtain the final result.
		Most prominent examples, where this is usually necessary, are 
		certain large-deviation simulation techniques.
		However, there are multiple approaches to do this, depending on the 
		data and individual requirements.
		Here, a software tool is presented, containing multiple algorithms, to 
		aid with this task.
		An introduction to the available methods is presented together with a 
		short tutorial using exemplary data.
	\end{abstract}	
	\maketitle
	\tableofcontents
	
	\section{Introduction}
		In large-deviation Monte-Carlo simulations 
		\cite{Hartmann_2002,Hartmann_2014,Boerjes_2019,Werner_2021} usually multiple runs are performed that are targeting different ranges for the 
		quantity of interest to obtain its distribution in the desired regime 
		of rare events.
		One important part of the data post-processing is to connect the 
		individual histograms to one final distribution.
		The software tool that is showcased here was originally developed with 
		this specific use case in mind, but also extends beyond.
		It can be useful for any variant of Monte-Carlo simulation, even 
		without	a relation to large-deviations, and experimental data alike.
		
		In the following, it is first explained how to obtain and run the tool.  
		Subsequently, there is a description of the implemented algorithms.
		Finally, a tutorial is given that demonstrates the core functionality 
		using exemplary data.
		\subsection{How to obtain the tool?}
		The tool is available through the repository \cite{git-repository},
		from where it can either be cloned via git or directly downloaded by a 
		web-browser.
		It comes in the form of a python3-script.
		\subsection{How to run the tool?}
		To run the script, a recent version of \textit{Anaconda} 
		\cite{anaconda} is recommended.
		Otherwise, the following libraries must be installed in a local 
		\textit{python3} environment:
		\textit{numpy}, \textit{gmpy2}, \textit{matplotlib} and \textit{scipy}.
		As a starting point, typing
		\begin{verbatim} 
		  python3 distribution_gluer.py --help
		\end{verbatim}
		into a command line will display an overview of all options.
		The displayed help-output also serves as the main reference for the
		program.

	\section{Gluing Algorithms}
		The tool can handle raw data, which is naively just a file containing
		one column of real valued numbers, or data that is already processed 
		into a histogram.
		For raw data, the tool can determine the corresponding histograms by 
		itself.	
		In any case, as a result of numerical simulations
		\cite{BPG_Hartmann_2015,Newman_Barkema_1999} or obtained otherwise, let 
		there be $N$ histograms $H_{\Theta_i}(x)$ with $i=1,\dots,N$ for an 
		arbitrary bin center value $x$. 
		The corresponding errors are denoted by $\sigma_{\Theta_i}(x)$.
		The value $\Theta_i$ is a simulation parameter on which the histograms 
		depend, usually the temperature or a temperature like parameter.
		For example in large deviation simulations
		\cite{Hartmann_2002,Hartmann_2014,Boerjes_2019,Werner_2021}, these are typically used to control 
		the extend by which the simulation is steered towards rare events.
		Here, $\Theta_i$ is referred to as a pseudo-temperature, even though it 
		can be an actual temperatures as well.
		This dependence can, but does not have to, result in a bias that needs 
		to be accounted for prior to connecting the distributions by 
		multiplying with a factor $f(\Theta, x)$:
		\begin{align}\label{eq:reweighted:histogram}
			H_i(x) &= H_{\Theta_i}(x) f(\Theta_i, x) \\
			\sigma_i(x) &= \sigma_{\Theta_i}(x) f(\Theta_i, x).
		\end{align}
		This factor could be just $f(\Theta, x) = 1$ in case all 
		pseudo-temperatures are the same $\Theta_i = \Theta$, but the 
		histograms were determined over different intervals (e.g. as it is the 
		case for Wang-Landau sampling \cite{Wang_Landau_2001, Vogel_2014}).
		Or when the overall quantity of interest is the energy-state-density of 
		a physical system following the Boltzmann-distribution, the factor 
		would be $f(\Theta, x) = \exp{(x/\Theta)}$.
		
		At this point in time, the tool offers two ways, namely the 
		Least-Squares (LSQRS) and the Ferrenberg-Swendsen (FS) 
		\cite{ferrenberg_1989} method, to connect histograms/distributions, 
		which are described briefly in the following.
		\subsection{Least-Squares-Method}
		The LSQRS method tries to minimize the squared difference
		\begin{equation}\label{eq:squared:difference}
			\Delta(g_i,g_j) =\sum_{x\in\mathcal{M}_{i,j}} \frac{(g_i H_i(x) - g_j H_j(x))^2}{\sigma_i^2(x) + \sigma_j^2(x)}
		\end{equation}
		between two histograms $i,j = 1,\dots, N$ and $i\neq j$, where 
		$\mathcal{M}_{i,j}$ is the set of bins $x$ for which both histograms 
		have a non-zero count, i.e. $H_i(x) \neq 0 \wedge H_j(x) \neq 0$.
		Additionally, the weights $g_i,g_j \neq 0$ have to fulfill the side 
		condition of a normalized total distribution.
		There is an analytical solution to eq. \eqref{eq:squared:difference} 
		(see appendix \ref{sec:lsqrs:solution}) resulting in a
		weighting factor $g_i$ for each histogram.
		Subsequently, these are used to determine the final probability density 
		and uncertainty
		\begin{align}\label{eq:probability:density}
			p(x) &=
			\frac{\sum_{i=1}^N \frac{g_i H_i(x)}{(g_i \sigma_i(x))^2}}{
			\sum_{j=1}^N \frac{1}{(g_j \sigma_j(x))^2}}\\
			\sigma_p(x) &= \frac{1}{\sqrt{\sum_{i=1}^N \frac{1}{(g_i \sigma_i(x))^2}}}.
		\end{align}
		
		\subsection{Ferrenberg-Swendsen-Method}
		The FS method assumes a particular bias of $f(\Theta_i, x) = 
		\exp{(x/\Theta_i)}$ on the histograms $H_{\Theta_i}(x)$, which is in 
		contrast to the LSQRS method that works with any form of bias.
		According to \cite{ferrenberg_1989}, the final distribution is obtained 
		by solving the equations
		\begin{align}\label{eq:fs:method}
			p(x) &= \frac{\sum_{i=1}^N H_{\Theta_i}(x)}{\sum_{j=1}^N e^{-\frac{x}{\Theta_j}+h_j}\sum_{\tilde{x}} H_{\Theta_j}(\tilde{x})} \\
			e^{-h_i} &= \sum_x p(x) e^{-\frac{x}{\Theta_i}}
		\end{align}
		numerically for $h_i$ ($i=1,\dots,N$) with an iterative procedure.
		The uncertainty on $p(x)$ is given by \cite{ferrenberg_1989}
		\begin{equation}\label{eq:fs:uncertainty}
			\sigma_p(x) = \frac{p(x)}{\sqrt{\left( \sum_{i=1}^N H_{\Theta_i}(x)\right)}}.
		\end{equation}
		
	\section{Tutorial}
		This tutorial assumes basic knowledge in using a command line 
		interpreter/shell (e.g. \textit{bash}).
		The tool comes with example data contained in the archive file 
		\textit{example\_data.zip},
		which should be unpacked before trying the following commands.
		The data can either be in raw form or already be histogramed.
		For reference on the specific file formatting, see the \textit{-f} 
		option of the help-output.
		In order to process raw data the \textit{-c} option has to be selected, 
		which will let the tool create the histograms by itself (see 
		\cite{numpy_binning} for technical reference; the \textit{bins='auto'} 
		argument is used).
		For convenience also the \textit{-P} option is set in the following 
		commands, yielding a plot of the result.
		The displayed figure shows in the bottom plot the original histograms 
		and in the top one, depending on the gluing method, for LSQRS the 
		individual reweighted histograms and for FS the final distribution.
		The legend indicates the corresponding pseudo-temperatures when
		provided.
		The final distribution is written to the output file 
		\textit{output.txt} as specified by the \textit{-o} option.
		If no gluing method is set, the tool defaults to LSQRS.
		
		\subsection{Raw unbiased data}
		In order to combine all files in the directory 
		\textit{raw\_unbiased\_data} into one distribution, the command
		\begin{verbatim}
		  python3 distribution_gluer.py -f \
		  example_data/raw_unbiased_data/* -c -P \
		  -o output.txt
		\end{verbatim}
		can be used.
		\begin{figure}
			\includegraphics[width=\linewidth]{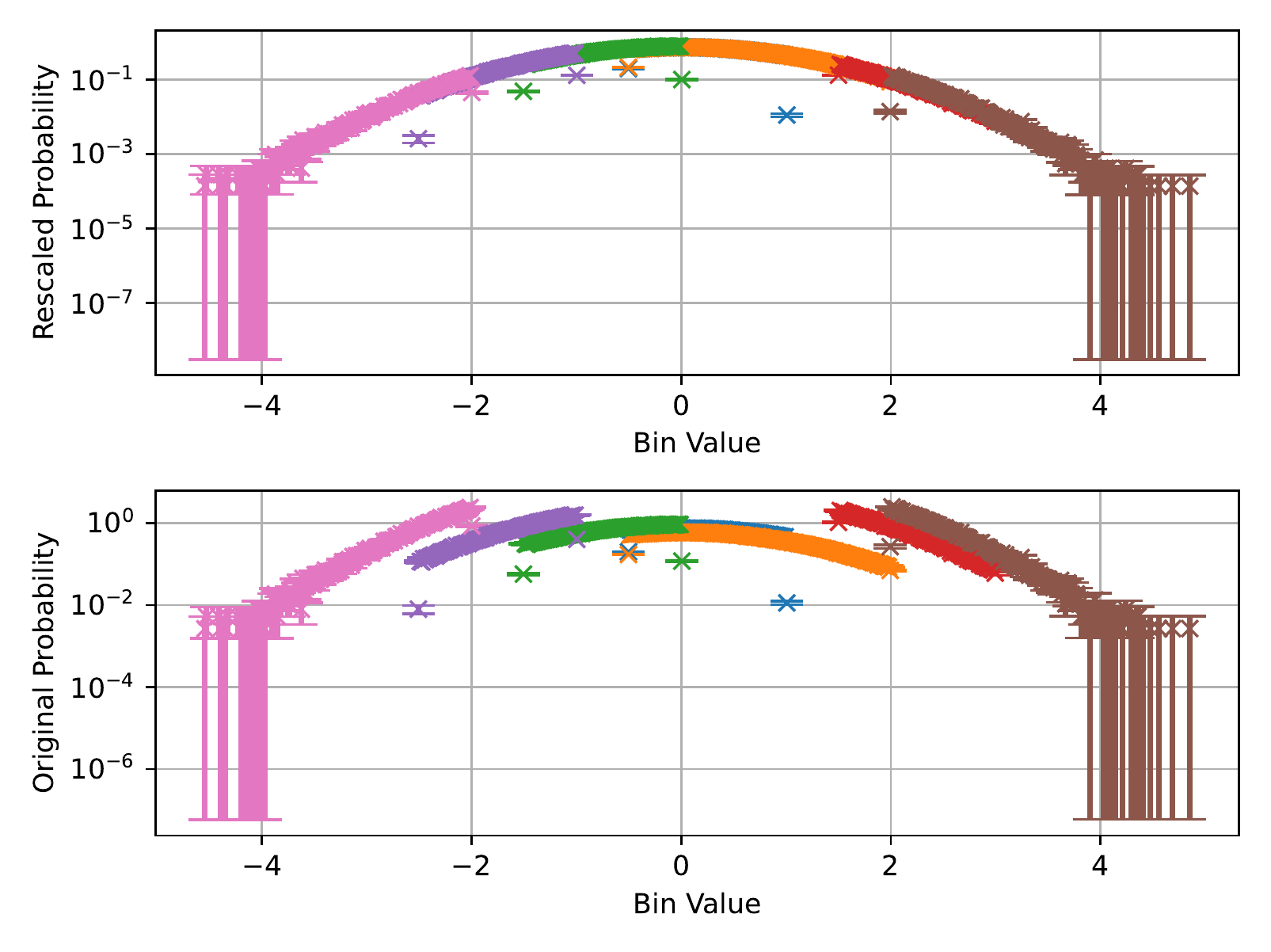}
			\caption{\label{fig:raw:data:gluing:example}Program output plot
				example for raw unbiased data. 
				Top: Glued/Rescaled distributions using the LSQRS method.
				Bottom: Original histograms determined by the gluing tool.}
		\end{figure}
		An example of how the displayed plot might look like is given in Fig. \ref{fig:raw:data:gluing:example}.\\
		To improve the result, separate binning intervals can be specified using
		the \textit{-b BINNING\_SEPERATOR1 BINNING\_SEPERATOR2 ...} option.
		Each of the separator variables marks the interval edges
		for which the internal binning algorithm is run independently.
		This can be helpful, when there are regions with fewer data points:
		\begin{verbatim}
		  python3 distribution_gluer.py -f \
		  example_data/raw_unbiased_data/* -c -P \
		  -o output.txt -b -3 3
		\end{verbatim}
		It is also possible to completely ignore any standard deviation on the
		histogram bins with the \textit{-i} option:
		\begin{verbatim}
				  python3 distribution_gluer.py -f \
				  example_data/raw_unbiased_data/* -c -P \
				  -o output.txt -b -3 3 -i
		\end{verbatim}
		Sometimes it is useful to cut off outlying values from the individual 
		histograms, before applying the LSQRS-method.
		The \textit{-s STD\_MULTIPLE} option will discard any bins that are 
		more than the specified multiple of the standard deviation away from 
		the mean: 
		\begin{verbatim}
		  python3 distribution_gluer.py -f \
		  example_data/raw_unbiased_data/* -c -P \
		  -o output.txt -b -3 3 -s 3
		\end{verbatim}
		\subsection{Histogrammed unbiased data}
		Like with the case of raw unbiased data, it is possible to combine data 
		that is already processed into histograms: 
		\begin{verbatim}
		  python3 distribution_gluer.py -f \
		  example_data/histogrammed_unbiased_data/* \
		  -P -o output.txt
		\end{verbatim}
		\subsection{Raw biased data}
		For biased data, pseudo temperatures have to be provided
		(option \textit{-p PSEUDO1 PSEUDO2 ...}), where a bias of the form
		$f(\Theta_i, x) = \exp{(x/\Theta_i)}$ is assumed.
		The FS method is selected via the \textit{-g FS} option. In the 
		following command, also a file containing specific bin edges (option 
		\textit{-B BINNING\_FILE}) is provided:
		\begin{verbatim}
		  PSEUDO={-10.0,10.0,-2.0,2.0,-4.0,4.0,\
		  	-6.0,6.0,-8.0,8.0,inf};\
		  python3 distribution_gluer.py -f \
		  $(eval echo example_data/raw_biased_data\
		  /pseudo_$PSEUDO.dat) \
		  -p $(eval echo $PSEUDO) \
		  -B example_data/raw_biased_data\
		  /bin_edges.txt -c -P -o output.txt -g FS
		\end{verbatim}
		\subsection{Histogrammed biased data}
		Again, it is also possible to work with histograms of biased data 
		directly:
		\begin{verbatim}
		  PSEUDO={-10.0,10.0,-2.0,2.0,-4.0,4.0\
		  ,-6.0,6.0,-8.0,8.0,inf};\
		  python3 distribution_gluer.py -f \
		  $(eval echo example_data\
		  /histogrammed_biased_data\
		  /pseudo_$PSEUDO.dat) \
		  -p $(eval echo $PSEUDO) -P \
		  -o output.txt -g FS
		\end{verbatim}
		\begin{figure}
			\includegraphics[width=\linewidth]{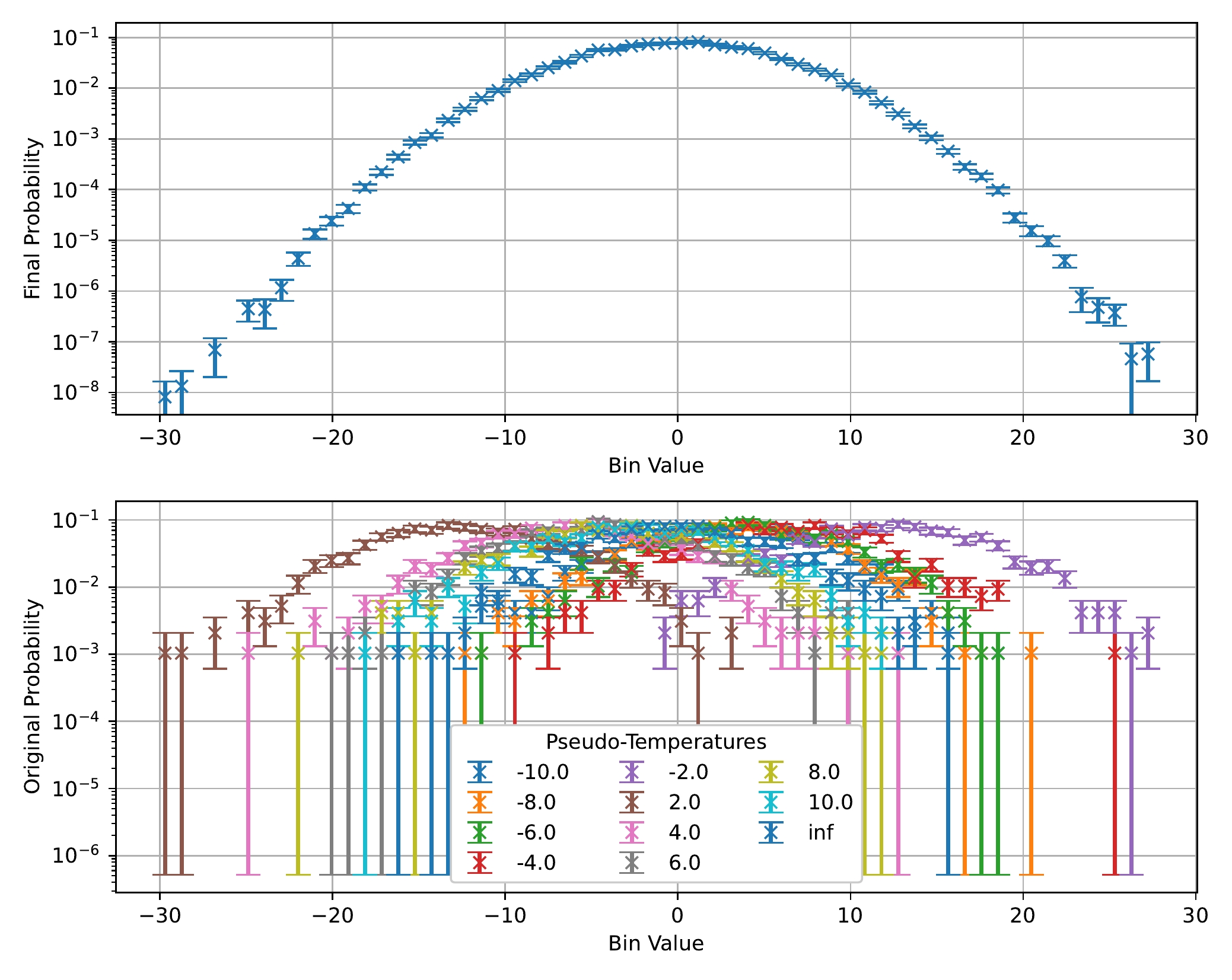}
			\caption{\label{fig:histogrammed:biased:data:gluing:example}Program 
				output plot example for biased histogrammed data. 
				Top: Glued final distribution using the FS method.
				Bottom: Original biased histograms at different 
				pseudo-temperatures $\Theta_i$ provided to the gluing tool.}
		\end{figure}
		An example of the program output plot for this command is displayed
		in Fig. \ref{fig:histogrammed:biased:data:gluing:example}.\\
		Biased histograms can also be treated with the default LSQRS method 
		using:
		\begin{verbatim}
		  PSEUDO={-10.0,10.0,-2.0,2.0,-4.0,4.0\
		  	,-6.0,6.0,-8.0,8.0,inf}; \
		  python3 distribution_gluer.py -f \
		  $(eval echo example_data\
		  /histogrammed_biased_data\
		  /pseudo_$PSEUDO.dat) \
		  -p $(eval echo $PSEUDO) -P -o output.txt
		\end{verbatim}
		
	\section{Final Remarks}
		The tool is in continuous development and any comments and suggestions	
		for improvement are very welcome.
		The author wishes to thank Alexander K. Hartmann for carefully reading 
		the manuscript and testing the program.

	\appendix
	\section{Analytical solution for LSQRS method weight factors} 
	\label{sec:lsqrs:solution}
		Rewriting eq. \eqref{eq:squared:difference} with relative weights
		$g_{ij} := g_j / g_i$ (relative weights have two indices and absolute 
		weights one) yields
		\begin{equation}\label{eq:squared:difference:with:ratio}
			\tilde{\Delta}(g_{ij}) := \sum_{x\in\mathcal{M}_{i,j}}
			\frac{(H_i(x) - g_{ij} H_j(x))^2}{\sigma_i^2(x) + \sigma_j^2(x)}.
		\end{equation}
		The minimal squared difference must satisfy the condition
		$\frac{\partial \tilde{\Delta}}{\partial g_{ij}} \overset{!}{=} 0$,
		resulting in
		\begin{equation}\label{eq:relative:weight:factor}
			g_{ij} = \frac{\sum_{x\in\mathcal{M}_{i,j}} \frac{H_i(x) 	H_j(x)}{\sigma_i^2(x) + \sigma_j^2(x)}}
			{\sum_{x'\in\mathcal{M}_{i,j}}\frac{H^2_j(x')}{\sigma_i^2(x') + 	\sigma_j^2(x')}}.
		\end{equation}
		For relative weights between non-overlapping histograms
		(i.e. $\mathcal{M}  = \emptyset$) eq. \eqref{eq:relative:weight:factor} 
		is not applicable.
		However, these can be calculated in a chain like manner.
		When $g_{ij}$ and $g_{jk}$ with $i \ne j \ne k$ are known and the 
		relative weight of interest $g_{ik}$ can not be calculated directly via 
		eq. \eqref{eq:relative:weight:factor}, it is still possible to use
		\begin{equation}\label{eq:relative:weight:factor:chain}
			g_{ik} = \frac{g_k}{g_i} = \frac{g_j}{g_i} \frac{g_k}{g_j}
			= g_{ij} g_{jk}.
		\end{equation}
		By fixing one of the absolute weight factors (e.g. $g_i$) such that the 
		total distribution is normalized, all other weights $g_j$ are determined by
		\begin{equation}\label{eq:absolute:weights:from:relative:weights}
			g_j = g_{ij} g_i.
		\end{equation}

	\bibliography{main}
		
\end{document}